\nonstopmode

\documentstyle[12pt,twoside]{article}

\newcommand{\beqn}{\begin{eqnarray}}
\newcommand{\eeqn}{\end{eqnarray}}
\newcommand{\be}{\begin{equation}}
\newcommand{\ee}{\end{equation}}
\newcommand{\ba}{\begin{array}}
\newcommand{\ea}{\end{array}}

\newcommand{\R}{{\rm\bf R}}
\newcommand{\C}{{\rm\bf C}}
\newcommand{\pa}{\partial}
\newcommand{\re}{\ref}
\newcommand{\ci}{\cite}
\newcommand{\la}{\label}
\newcommand{\bfr}{\begin{flushright}}
\newcommand{\efr}{\end{flushright}}
\newcommand{\bfl}{\begin{flushleft}}
\newcommand{\efl}{\end{flushleft}}
\newcommand{\fr}{\frac}

\newcommand{\ti}{\tilde}
\newcommand{\si}{\sigma}
\newcommand{\Si}{\Sigma}
\newcommand{\al}{\alpha}

\newcommand{\ds}{\displaystyle}
\newcommand{\ve}{\varepsilon}

\newcommand{\de}{\delta}
\newcommand{\De}{\Delta}

\newcommand{\om}{\omega}
\newcommand{\Om}{\Omega}

\newcommand{\br}{|\kern-.25em|\kern-.25em|}
\newcommand{\brr}{{|\kern-.15em|\kern-.15em|\kern-.15em}\,}

\newtheorem{theorem}{Theorem}[section]

\newtheorem{definition}[theorem]{Definition}

\newtheorem{lemma}[theorem]{Lemma}

\newtheorem{remark}[theorem]{Remark}

\newtheorem{cor}[theorem]{Corollary}
\newtheorem{pro}[theorem]{Proposition}
\def\N{{\rm I\kern-.1567em N}}                              
\def\R{{\rm I\kern-.1567em R}}                              
\def\C{{\rm C\kern-4.7pt                                    
\vrule height 7.7pt width 0.4pt depth -0.5pt \phantom {.}}\,}
\def\Z{{\sf Z\kern-4.5pt Z}}                                
\def\Re {{\rm Re\, }}                                       
\def\Im {{\rm Im\,}}                                        

\newcommand{\supp}{\mathop{\rm supp}\nolimits}

\newcommand{\bo}{{\hfill\loota}}
\newcommand{\loota}{\hbox{\enspace{\vrule height 7pt depth 0pt width
      7pt}}}

\begin{document}
\begin{titlepage}
\begin{center}
{\Large\bf Convergence to equilibrium distribution.\\
~\\
 The Klein-Gordon equation coupled to a particle }\\
\vspace{1cm}
{\large T.V. Dudnikova}
\footnote{Partially supported by the research grant 
of RFBR (06-01-00086)}\\
{\small\it Elektrostal Polytechnical Institute\\
 Elektrostal 144000, Russia\\
e-mail:~dudnik@elsite.ru}
\end{center}
 \vspace{2cm}

 \begin{abstract}
We consider the Hamiltonian system consisting
of a Klein-Gordon vector field and a particle in $\R^3$.  
The initial date of the system is a random function
with a finite mean density of energy which also satisfies
 a Rosenblatt- or Ibragimov-type mixing condition.
Moreover, initial correlation functions are translation-invariant.
We study the distribution $\mu_t$ of the solution at time $t\in\R$.
The main result is the convergence of $\mu_t$ to a Gaussian 
measure as $t\to\infty$, where $\mu_\infty$ is translation-invariant.


 \end{abstract}
\end{titlepage}

\section{Introduction}
The paper concerns  problems of long-time convergence to an
 equilibrium distribution in a coupled system.
 We have proved the convergence for partial differential 
equations of hyperbolic type in $\R^n$, $n\ge2$, 
in \cite{DKKS,DKRS,DKS,DKM2}; for harmonic crystals in \ci{DKS1, DKM1}, 
and for a scalar field coupled to a harmonic crystal in \ci{DK}.
Here we treat a particle coupled to a Klein-Gordon  or wave vector equation.

Let us outline our main result and the strategy of the proof.
(For the formal definitions and statements, see Section 2.)
Consider the  Hamiltonian system 
consisting of a real-valued vector field $\varphi(x)$, $x\in\R^3$, 
and a particle with position $q\in\R^3$. The Hamiltonian functional is
\beqn\la{Ham}
H(\varphi,q,\pi,p)&=&
\sum\limits_{n=1}^{d}
\int\limits_{\R^3}\Bigr(\frac{|\nabla\varphi_n(x)|^2}{2}+
\frac{m_n^2|\varphi_n(x)|^2}{2}+\frac{|\pi_n(x)|^2}{2}
+\varphi_n(x)q\cdot\nabla\rho_n(x)\Bigr)\,dx\nonumber\\
&&+\frac{1}{2}\Big(|p|^2+\omega^2 |q|^2\Big).
\eeqn
Here $m_n\ge0$, $\omega>0$, 
$\varphi(x)=(\varphi_1(x),\dots,\varphi_d(x))\in\R^d$
and correspondingly for $\pi(x)$, "$\cdot$" stands for the scalar product
in the Euclidean space $\R^3$.
 Taking formally  variational derivatives in (\ref{Ham}),
the coupled dynamics becomes
\beqn
\la{1}
&\dot\varphi_n(x,t)=\pi_n(x,t),&
\dot\pi_n(x,t)=(\triangle-m_n^2)\varphi_n(x,t)-q(t)\cdot\nabla\rho_n(x),\,\,\,
n=1,\dots,d,\\
&\dot q(t)=p(t),&
\dot p(t)=-\omega^2 q(t)-\sum\limits_{n=1}^{d}
\int\varphi_n(x,t)\nabla\rho_n(x)\,dx,\,\,\,\,t\in\R.\la{4}
\eeqn
To state our main results we formulate some assumptions on a 
constant $\omega>0$ and a 
coupled function $\rho(x)=(\rho_1(x),\dots,\rho_d(x))$.
\\
{\bf A1}. The matrix $(\om^2-m_*^2) I-K$ is positive definite,
where $m_*=0$ if $m=(m_1,\dots,m_d)=0$ and
$m_*=\min\{m_n:m_n\not=0,n=1,\dots,d\}$ if $m\not=0$,
 $K=(K_{ij})_{i,j=1}^3$ is the $3\times3$ matrix with
the matrix elements $K_{ij}$,
$$
K_{ij}=\sum\limits_{n=1}^d\frac{1}{(2\pi)^3}
\int\frac{k_ik_j|\hat\rho_n(k)|^2}{k^2+m_n^2-m_*^2}\,dk.
$$
{\bf A2}. The function $\rho(x)$ is a vector  real-valued smooth function,
$\rho(-x)=\rho(x)$, $\rho(x)=0$ for $|x|\ge R_{\rho}$.\\
{\bf A3}. For all  $n=1,\dots,d$ and $k\in\R^3\setminus\{0\}$,
$$
\hat\rho_n(k)=\int e^{ikx}\rho_n(x)\,dx\not=0. 
$$
This assumption can be weakened (see Remark \ref{r7.6}).

We study the Cauchy problem for the system (\ref{1})--(\ref{4})
with initial conditions
\be\la{5}
\varphi(x,0)=\varphi^0(x),~~\pi(x,0)=\pi^0(x),~~
q(0)=q^0,~~p(0)=p^0.
\ee
Let  us write 
$Y_0\equiv(\varphi^0(x),q^0,\pi^0(x),p^0)$,
$Y(t)\equiv(\varphi(x,t),q(t),\pi(x,t),p(t))$.
Then  the system (\ref{1})--(\ref{5}) writes as
\be\la{1.1'}
\dot Y(t)={\cal F}(Y(t)),\,\,\,t\in\R;\,\,\,\,Y(0)=Y_0.
\ee
We assume that the initial date $Y_0$ is a
random element of a real functional space ${\cal E}$
consisting of states with a finite local energy,
see Definition \ref{d1.1} below.
  The distribution of $Y_0$ is a probability measure  $\mu_0$
of mean zero satisfying conditions {\bf S1}--{\bf S3}.
In particular, the correlation functions of the measure  $\mu_0$  
is translation-invariant.
For a given $t\in\R$, we denote by $\mu_t$ the probability measure
on  ${\cal E}$ defining the distribution of the solution $Y(t)$ to the 
problem (\re{1.1'}).

Our main result gives  the weak convergence of the measures
$\mu_t$  to a limit measure $\mu_{\infty}$,
\be\la{1.8i}
\mu_t \rightharpoondown
\mu_\infty,\,\,\,\, t\to \infty.
\ee
The measure $\mu_{\infty}$ is 
 a translation-invariant Gaussian measure on ${\cal E}$.
 Similar results hold as $t\to-\infty$ because the dynamics is time-reversible.
Moreover, in Section \ref{s.conv} we prove the convergence of the correlation
functions of the measures $\mu_t$ to a limit as $t\to\infty$.

We prove the convergence (\re{1.8i}) by using the strategy of
\ci{DKKS, DKRS} in two steps:\\
{\bf I.} The family of measures $\mu_t$, $t\geq 0$, is weakly
compact in an appropriate Fr\'echet space.\\
{\bf II.}
The characteristic functionals converge to a Gaussian functional,
\be\la{2.6i}
 \hat\mu_t(Z) = \int \exp({i\langle Y,Z\rangle})\,\mu_t(dY)
\rightarrow \ds \exp\{-\fr{1}{2}{\cal Q}_\infty (Z,Z)\},
\,\,\,\,t\to\infty,
\ee
where $Z$ is an arbitrary element of the dual space and
${\cal Q}_\infty$ is the quadratic form defined by (\ref{Qmu}),
$\langle \cdot,\cdot\rangle$ stands for the inner product
in a real Hilbert space $L^2(\R^3)\otimes\R^N$ with different $N=1,2,\dots$.

Let us explain the main idea of the proof.
At first we derive the decay of the order $(1+|t|)^{-3/2}$
(and the exponential decay in the case when $m=0$) 
for the local energy
of the solution $Y(t)$ to (\ref{1.1'}) assuming 
that the initial date $Y_0$ has a compact support 
(see Proposition \ref{l5.1}). 
Then we apply the integral representation (\ref{6.4})
of $Y(t)$ and prove a uniform bound (\ref{7.1.1}) 
for the mean local energy with respect to the measure 
 $\mu_t$, $t\ge0$.
Finally property {\bf I} follows from the Prokhorov Theorem.

To prove of {\bf II} we derive the asymptotic behavior of the solution
$Y(t)$ (see Proposition \ref{l7.1}), namely,
\be\la{0.1}
\langle Y(t),Z\rangle\sim
\sum\limits_{n=1}^d\langle W_n(t)(\varphi_n^0,\pi_n^0),\psi_n^Z\rangle,
\,\,\,t\to\infty,
\ee 
where $W_n(t)$ is a solving operator to the Cauchy problem (\ref{freeKG})
for the free wave or Klein -- Gordon equation, 
the functions $\psi_n^Z$ are expressed by $Z$ 
(see formula (\ref{hn})).
Then we apply the results of \cite{DKKS, DKRS}, where the weak convergence 
of the statistical solutions 
is proved for free wave and Klein-Gordon equations. 
 
\setcounter{equation}{0}
\section{Main results}
\subsection{Notation}
We assume that the initial data $Y_0$
 are given by an element of the real phase space ${\cal E}$ defined below.
\begin{definition}                 \la{d1.1'}
Let $ {\cal H} \equiv H_{loc}^1(\R^3)\oplus H_{loc}^0(\R^3)$
 be the Fr\'echet space
 of pairs $\phi\equiv(\varphi(x),\pi(x))$
with $\R^d$-valued functions $\varphi(x)$ and $\pi(x)$,
 which is endowed with the local energy seminorms
 \beqn \la{1.5}
\Vert \phi\Vert^2_{R}= \int\limits_{|x|<R}
(|\varphi(x)|^2+|\nabla\varphi(x)|^2+|\pi(x)|^2) dx<\infty,
~~\forall R>0.
 \eeqn
  \end{definition}
\begin{definition}\la{d1.1}
Let ${\cal E} \equiv {\cal H}\oplus\R^3\oplus\R^3$
be the Fr\'echet space of vectors $Y\equiv(\phi(x),q,p)$,
with the local energy seminorms
\be\la{2.1'}
\Vert Y\Vert^2_{{\cal E},R}=
\Vert\phi\Vert^2_{R}+|q|^2+|p|^2, ~~\forall R>0.
\ee
\end{definition}

Now we formulate the following condition on $\rho$ (cf. condition {\bf A1}).\\
{\bf A1'}. The matrix $\om^2 I-K_0$ is positive definite,
where $K_0=(K_{0,ij})_{i,j=1}^3$ is the $3\times3$ matrix with
the matrix elements $K_{0,ij}$,
$$
K_{0,ij}=\sum\limits_{n=1}^d\frac{1}{(2\pi)^3}
\int\frac{k_ik_j|\hat\rho_n(k)|^2}{k^2+m_n^2}\,dk.
$$
Note that if $m=0$ the conditions {\bf A1} and {\bf A1'} coincide.
If $m\not=0$, condition {\bf A1'} is weaker than {\bf A1}.
 \begin{pro}    \la{p1.1}
Let conditions {\bf A1'} and {\bf A2} hold. Then 
(i) for every $Y_0 \in {\cal E}$, the Cauchy problem (\re{1.1'})
has a unique solution $Y(t)\in C(\R, {\cal E})$.\\
 (ii) For every $t\in \R$, the operator $U(t):Y_0\mapsto  Y(t)$
 is continuous on ${\cal E}$.  Moreover, for every $R>R_\rho$, $T>0$,
\be\la{3.2.1}
\sup\limits_{|t|\le T}
\Vert U(t) Y_0\Vert^2_{{\cal E},R}\le
C(T)\Vert Y_0\Vert^2_{{\cal E},R+T}.
\ee
\end{pro}
Proposition \ref{p1.1} is proved in Section 3.
\medskip

Let us choose a function $\zeta(x)\in C_0^\infty(\R^3)$ with $\zeta(0)\ne 0$.
Denote by $H^s_{\rm loc}(\R^3)$, $s\in \R$,  the local Sobolev spaces
of $\R^d$-valued functions $\varphi$,
i.e., the Fr\'echet spaces
of distributions $\varphi\in D'(\R^3)$ with the finite seminorms
$
\Vert \varphi\Vert_{s,R}
:=\Vert\Lambda^s\Big(\zeta(x/R)\varphi\Big)\Vert_{L^2(\R^3)},
$
where $\Lambda^s \psi:=F^{-1}_{k\to x}(\langle  k\rangle^s\hat \psi(k))$,
$\langle  k\rangle:=\sqrt{|k|^2+1}$, and $\hat \psi$ is the Fourier
transform of a tempered distribution $\psi$.
\begin{definition}\la{d1.2}
For $s\in\R$  denote
${\cal H}^{s}\equiv H_{\rm loc}^{1+s }(\R^3)
\oplus H_{\rm loc}^{s}(\R^3)$,
${\cal E}^{s}\equiv {\cal H}^{s}\oplus\R^3\oplus\R^3$.
\end{definition}

Using standard techniques of pseudodifferential operators
and Sobolev's Theorem (see, e.g., \ci{H3}), it is possible to prove that
 ${\cal E}^0={\cal E}\subset {\cal E}^{-\ve }$ for every $\ve>0$,
and the embedding  is compact.

\subsection{Random solution. Convergence to equilibrium}

Let $(\Om,\Sigma,P)$ be a probability space
 with expectation $E$
and ${\cal B}({\cal E})$ denote the Borel $\si$-algebra
in ${\cal E}$.
We assume that $Y_0=Y_0(\om,x)$ in (\re{1.1'})
is a measurable random function
with values in $({\cal E},\,{\cal B}({\cal E}))$.
In other words, $(\om,x)\mapsto Y_0(\om,x)$ is a measurable map
$\Om\times\R^3\to\R^{2d+6}$
 with respect to the (completed) $\si$-algebra
$\Sigma\times{\cal B}(\R^3)$ and ${\cal B}(\R^{2d+6})$.
Then $Y(t)=U(t) Y_0$ is also a measurable random function with values in
$({\cal E},{\cal B}({\cal E}))$ owing to Proposition \re{p1.1}.
We denote by $\mu_0(dY_0)$ a Borel probability measure in ${\cal E}$
giving the distribution of  $Y_0$.
Without loss of generality, we assume 
$(\Om,\Si,P)=({\cal E},{\cal B}({\cal E}),\mu_0)$
and $Y_0(\om,x)=\om(x)$ for
$\mu_0(d\om)\times dx$-almost all $(\om,x)\in{\cal E}\times\R^3$.
\begin{definition}
$\mu_t$ is a Borel probability measure in ${\cal E}$
which gives the distribution of $Y(t)$:
\begin{eqnarray}\nonumber
\mu_t(B) = \mu_0(U(-t)B),\,\,\,\,
\forall B\in {\cal B}({\cal E}), \,\,\,   t\in \R.
\eeqn
\end{definition}

Our main objective is to prove
 the weak convergence of the measures $\mu_t$
in the Fr\'echet spaces ${\cal E}^{-\ve }$ for each  $\ve>0$,
\beqn\la{1.8}
\mu_t\,\buildrel {\hspace{2mm}{\cal E}^{-\ve }}\over
{- \hspace{-2mm} \rightharpoondown }
\, \mu_\infty  \quad{\rm as}\quad t\to \infty,
\eeqn
where $\mu_\infty$ is a limit measure on ${\cal E}$.
 This means the convergence
 \beqn\nonumber
 \int f(Y)\mu_t(dY)\rightarrow
 \int f(Y)\mu_\infty(dY)\quad{\rm as}\quad t\to \infty
 \eeqn
 for any bounded continuous functional $f(Y)$
 on  ${\cal E}^{-\ve }$.

\begin{definition}
The correlation functions of measure $\mu_t$ are
defined by
\beqn\nonumber
Q_t(x,y)\equiv E \Big(Y(x,t)\otimes Y(y,t)\Big),~~~~
{\rm for~~almost~~all}\,\,\,\, x,y\in\R^3,
\eeqn
if the expectations in the right hand side are finite.
\end{definition}

We set ${\cal D}={\cal D}_0\oplus\R^3\oplus\R^3$,
${\cal D}_0:=[C_0^{\infty}(\R^3)\otimes\R^d]^2$, and
$$
\langle Y,Z\rangle:=\langle \phi,\psi\rangle+q\cdot u+p\cdot v
$$
for $Y=(\phi,q,p)\in {\cal E}$, and 
$Z=(\psi,u,v)\in  {\cal D}$.
For a  probability  measure $\mu$ on  ${\cal E}$
 denote by $\hat\mu$ the characteristic functional (Fourier transform)
$$
\hat\mu(Z)\equiv\int\exp(i\langle Y,Z\rangle )\,\mu(dY),\,\,\,
 Z\in {\cal D}.
$$
A  measure $\mu$ is called Gaussian (with zero expectation) if
its characteristic functional has the form
$$
\ds\hat {\mu} (Z)=\ds \exp\{-\fr{1}{2}
 {\cal Q}(Z,Z)\},\,\,\,Z \in {\cal D},
$$
where ${\cal Q}$ is a  real nonnegative quadratic form in ${\cal D}$.
A measure $\mu$ is called translation-invariant if
$$
\mu(T_h B)= \mu(B),\,\,\,\,\,\forall B\in{\cal B}({\cal E}),
\,\,\,\, h\in\R^3,
$$
where $T_h Y(x)= Y(x-h)$.

\subsection{Main theorem}

We assume that the initial measure $\mu_0$
has the following properties {\bf S0}--{\bf S3}.\\
{\bf S0} $\mu_0$ has zero expectation value,
$EY_0(x)\equiv\ds\int Y_0(x)\,\mu_0(dY_0)= 0$, $x\in\R^3$.
\smallskip\\
{\bf S1} $\mu_0$ has finite mean energy density, i.e., 
\be\la{med}
E\Big(|\varphi^0(x)|^2+|\nabla\varphi^0(x)|^2
+|\pi^0(x)|^2+|q^0|^2+|p^0|^2\Big) <\infty.
\ee

Denote by $\nu_0:=P\mu_0$, where $P:(\phi^0,q^0,p^0)\in {\cal E}
\to\phi^0\in{\cal H}$.\\
{\bf S2} The correlation functions of the measure $\nu_0$ 
are translation invariant, i.e., for $n,n'\in\bar d=\{1,\dots,d\}$,
 $x,y\in\R^3$,
\beqn\la{2.4'}
\ba{rcl}
&E\Big(\varphi^0_n(x)\otimes\varphi^0_{n'}(y)\Big):=
q^{00}_{0,nn'}(x-y),&
E\Big(\varphi^0_n(x)\otimes\pi^0_{n'}(y)\Big):=
q^{01}_{0,nn'}(x-y),\\
&E\Big(\pi^0_n(x)\otimes\varphi^0_{n'}(y)\Big):=
q^{10}_{0,nn'}(x-y),&
E\Big(\pi^0_n(x)\otimes\pi^0_{n'}(y)\Big):=
q^{11}_{0,nn'}(x-y).
\ea
\eeqn 

Now we formulate the mixing condition for the measure $\nu_0$.

 Let $O(r)$ be the set of all pairs of open convex subsets
 $ {\cal A}, {\cal B} \subset \R^3$ at the distance
not less than $r$,
 $d({\cal A},{\cal B})\geq r$,  and let $\sigma ({\cal A})$
be the $\sigma $-algebra  in ${\cal H}$ generated by the
linear functionals $\phi\mapsto\, \langle \phi,\psi\rangle$,
where  $\psi\in {\cal D}_0$ with $\supp\psi\subset {\cal A}$.
Define the Ibragimov mixing coefficient
of a probability measure $\nu_0$ on ${\cal H}$
by the rule (cf \ci[Def. 17.2.2]{IL})
\beqn\la{ilc}
\varphi(r)\equiv
\sup_{({\cal A},{\cal B})\in O(r)} \sup_{
\ba{c} A\in\si({\cal A}),B\in\si({\cal B})\\ \nu_0(B)>0\ea}
\fr{| \nu_0(A\cap B) - \nu_0(A)\nu_0(B)|}{ \nu_0(B)}.
\eeqn
\begin{definition}\la{dmix}
 We say that the measure $\nu_0$ satisfies 
the strong uniform Ibragimov mixing condition if
$\varphi(r)\to 0$ as $r\to\infty$.
\end{definition}
{\bf S3} The measure $\nu_0$  satisfies the strong uniform
Ibragimov  mixing condition, and
\be\la{1.12}
\int\limits_0^{+\infty} r^2\varphi^{1/2}(r)dr <\infty.
\ee
 
Consider the following Cauchy problem
\be\la{freeKG}
\left\{\ba{l}
\ddot \varphi(x,t) =(\Delta-m_n^2)\varphi(x,t),\,\,\,\,t\in\R,\\
\varphi(x,t)|_{t=0}=\varphi^0(x),\,\,\,
\dot\varphi(x,t)|_{t=0}=\pi^0(x),\,\,\,\,x\in\R^3,
\ea\right.
\ee
where $m_n\ge0$, $\varphi(x,t)\in\R^1$.
The following lemma is proved in \cite[p.7]{DKKS}, \cite[p.1225]{DKRS}.
\begin{lemma} 
(i) For any $\phi^0=(\varphi^0,\pi^0)\in {\cal H}_1\equiv 
H^1_{loc}(\R^3)\oplus L^2_{loc}(\R^3)$, 
there exists a unique solution 
$\phi(t)=(\varphi(x,t),\dot\varphi(x,t))
\in C(\R, {\cal H}_1)$ to the Cauchy problem (\re{freeKG}).\\
 (ii) For any $t\in \R$, the operator $W_n(t):\phi^0\mapsto\phi(t)$
 is continuous in ${\cal H}_1$.
\end{lemma}

We define the operator $W(t)$ 
on the space ${\cal H}=[{\cal H}_1]^d$ by the rule
\be\la{W(t)}
W(t)(\phi^0_1,\dots,\phi^0_d)=(W_1(t)\phi^0_1,\dots,W_d(t)\phi^0_d).
\ee
Let ${\cal E}_n(x)$ be the fundamental solution of the 
operator  $-\De+m_n^2$, $n\in\bar d$. 
For almost all $x,y\in\R^3$, introduce the matrix-valued function 
$Q^\nu_{\infty}(x,y)= \left(q_{\infty,nn'}(x-y)\right)_{n,n'=1}^d$, where
\beqn\la{1.13}
q_{\infty,nn'}=\left(q^{ij}_{\infty,nn'}\right)_{i,j=0}^1=
\chi_{nn'}\frac{1}{2}\left(\ba{ll}
q_{0,nn'}^{00}+{\cal E}_n * q_{0,nn'}^{11} &
q_{0,nn'}^{01}-~~q_{0,nn'}^{10} \\
~\\
q_{0,nn'}^{10}-~~q_{0,nn'}^{01} & 
q_{0,nn'}^{11}+(-\De+m_n^2) q_{0,nn'}^{00}\ea\right),
\eeqn
where $\chi_{nn'}=1$ if $m_n=m_{n'}$, and $\chi_{nn'}=0$ otherwise,
 the functions $q^{ij}_{0,nn'}$, $i,j=0,1$, 
are defined in (\ref{2.4'}); 
and $*$ stands for the convolution of distributions.
\begin{remark}\la{r2.9}
{\rm According to \cite[Lemma 17.2.3]{IL}, the derivatives 
$\pa^\alpha q_{0,nn'}^{ij}$ are bounded by mixing coefficient:
$\forall\alpha\in\Z^3_+$ with $|\alpha|\le 2-i-j$
(including $\alpha=0$), $i,j=0,1$,
$$
|\pa^\alpha q_{0,nn'}^{ij}(z)|\le C\varphi^{1/2}(|z|),\quad \forall z\in\R^3,
\quad n, n'\in\bar d.
$$
Therefore, $\pa^\alpha q_{0,nn'}^{ij}\in L^p(\R^3)$, 
$p\ge1$ (see \cite[p.16]{DKKS}).
Hence, $q_{\infty,nn'}^{ij}\in L^1(\R^3)$ if $m_n\not=0$ by 
the bound (\ref{1.13}).
If $m_n=0$, (\ref{1.12}) implies the existence of the convolution
 ${\cal E}_n * q_{0,nn'}^{11}$.}
\end{remark}

Denote by
${\cal Q}^\nu_{\infty} (\psi,\psi)$ a real quadratic form on
 ${\cal S}_0\equiv [S(\R^3)\otimes\R^d]^{2}$  defined by
\beqn\la{qpp}
{\cal Q}^\nu_{\infty} (\psi,\psi)=
\langle Q^\nu_{\infty}(x,y),\psi(x)\otimes \psi(y)\rangle
= \sum\limits_{n,n'=1}^d
\langle q_{\infty,nn'}(x-y),\psi_n(x)\otimes \psi_{n'}(y)\rangle.
\eeqn 
The following result can be proved by an easy adaptation
of the proof of  Theorem~B of \cite{DKKS, DKRS},
where the result is proved in the case $d=1$.
\begin{theorem} \la{l6.2}  
Let conditions {\bf S0}--{\bf S3} hold. Then\\
(i) the measures $\nu_{t}\equiv W(t)^*\nu_0$
weakly converge as $t\to\infty$ on the space ${\cal H}^{-\ve}$
for each $\ve>0$.\\
(ii) The limit measure  $\nu_{\infty}$ 
is a translation-invariant Gaussian measure on ${\cal H}$.\\
(iii) The characteristic functional of $\nu_{\infty}$ is of the form 
$$
\hat\nu_{\infty}(\psi)=\exp\Big\{-\frac{1}{2} 
{\cal Q}^\nu_{\infty}(\psi,\psi)\Big\},\,\,\,\, \psi\in {\cal S}_0.
$$
\end{theorem}
  
Let $Z=(\psi,u,v)\in{\cal D}$, i.e., 
$\psi=(\psi_1,\dots,\psi_d)\in {\cal D}_0$, $(u,v)\in \R^3\times\R^3$.
Denote
\be\la{hn}
\psi^Z=(\psi^Z_1,\dots,\psi^Z_d),\,\,\,\psi_k^Z:=\psi_k(x)-
\sum\limits_{n=1}^d\theta_{kn}(x)
+\alpha_k(x)\cdot  u+\beta_k(x)\cdot v,
\,\,\,k\in\bar d.
\ee
Here $\alpha_k=(\alpha^1_k,\alpha^2_k,\alpha^3_k)$, 
$\beta_k=(\beta^1_k,\beta^2_k,\beta^3_k)$,
\beqn
\alpha_k^i\equiv\alpha_k^i(x)
&=& -\ds\sum\limits_{r=1}^d\int\limits_0^{+\infty}
{\cal N}_{ir}(s)W'_k(-s)\left(\ba{c}
\nabla_r \rho_k(x)\\0\ea\right)\,ds,\la{al}\\
\beta_k^i\equiv\beta_k^i(x)
&=& -\ds\sum\limits_{r=1}^d\int\limits_0^{+\infty}
{\cal N}_{ir}(s)W'_k(-s)\left(\ba{c}
0\\\nabla_r \rho_k(x)\ea\right)\,ds,\la{beta}
\eeqn
\be\la{teta}
\theta_{kn}(x):=\sum\limits_{i=1}^3
\int\limits_0^{+\infty}
W'_k(-s)\alpha^i_k(x)
\left\langle W_n(s) {\cal R}_{in},\psi_n\right\rangle\,ds,\,\,\,\,
{\cal R}_{in}:=\left( \ba{ll} 0\\ \nabla_i\rho_n\ea \right),
\ee 
the matrix ${\cal N}(s)=({\cal N}_{ir}(s))_{i,r=1}^3$ is defined 
in (\ref{4.15}),
the operator $W'_n(t)$ is adjoint to the operator $W_n(t)$:
$$
\langle \phi, W'_n(t)\psi\rangle=\langle W_n(t)\phi,\psi\rangle,\quad
\psi\in [S(\R^3)]^2,\,\,\,\phi\in{\cal H}_1,\,\,\,\,t\in\R.
$$

Denote by ${\cal Q}_{\infty} (Z,Z)$ 
a real quadratic form in ${\cal D}$ of the form
\beqn\la{Qmu}
{\cal Q}_{\infty} (Z,Z)=
{\cal Q}^\nu_{\infty} (\psi^Z,\psi^Z),
\eeqn
where $\psi^Z$ is defined in (\ref{hn}).
Our main result is the following theorem.
\begin{theorem}\la{tA}
Let conditions {\bf A1}--{\bf A3} and {\bf S0}--{\bf S3} hold. Then\\
(i) the convergence in (\ref{1.8}) holds for any $\ve>0$. \\
(ii)  The limit measure
$ \mu_\infty $ is a Gaussian equilibrium
measure on ${\cal E}$.\\
(iii) The limit characteristic functional has the form
$$
\ds\hat { \mu}_\infty (Z)=\ds \exp\{-\fr{1}{2}
{\cal  Q}_\infty(Z,Z)\},~~ \,\,\, Z \in  {\cal D}.
$$
(iv) The measure $\mu_\infty$ is invariant, i.e.,
$U(t)^* \mu_\infty=\mu_\infty,\,\,\,\,t\in\R$.
\end{theorem}
\begin{remark}
{\rm Instead of the {\it strong uniform} Ibragimov mixing condition, 
it suffices to assume the {\it uniform} Rosenblatt mixing condition \cite{Ros} 
together with a higher degree ($>2$) in the bound (\re{med}), i.e., 
to assume that there exists  a $\delta$, $\de >0$, such that   
$$   
E\left(|\varphi^0(x)|^{2+\de}+|\nabla\varphi^0(x)|^{2+\de}   
+|\pi^0(x)|^{2+\de}+|q^0|^2+|p^0|^2\right) <\infty.   
$$
In this case, the condition (\re{1.12}) needs the following modification:   
$\ds\int_0^{+\infty} \ds r^2\al^{p}(r)dr <\infty$, 
where  $p=\min(\de/(2+\de), 1/2)$, 
$\al(r)$ is the  Rosenblatt mixing coefficient
 defined as in  (\re{ilc}) but without $\nu_0(B)$ in the denominator.   
}\end{remark}

\setcounter{equation}{0}
\section{Existence of solutions, a priori estimates}
In this section we prove Proposition \ref{p1.1}
by the similar arguments as in \cite[Lemma 6.3]{KSK}.

Let us represent the solution $Y(t)$ as the pair of the functions
$(Y^0(t),Y^1(t))$, 
where $Y^0(t)=(\varphi(t),q(t))$, $Y^1(t)=(\pi(t),p(t))$.

Denote by $H^s(\R^3)$ the Sobolev space of $\R^d$-valued functions.
Let $E$ be the Hilbert space of pairs $Y=(Y^0, Y^1)$, where
$Y^0=(\varphi,q)\in H^1(\R^3)\oplus\R^3$,
$Y^1=(\pi,p)\in H^0(\R^3)\oplus\R^3$,
with the finite norm
$
\Vert Y\Vert_E^2:=\ds\sum\limits_{n=1}^d\left(
\Vert \nabla \varphi_n\Vert^2+m_n^2
\Vert\varphi_n\Vert^2+\Vert\pi_n\Vert^2\right)+|q|^2+|p|^2.
$
Here $\Vert\cdot\Vert$ stands for the norm in $H^0(\R^3)$.
Now we prove the auxilary lemma.
\begin{lemma}\la{l3.1}
Let conditions {\bf A1'} and {\bf A2} hold. Then\\
(i) for every $Y_0\in E$, the Cauchy problem (\re{1.1'})
has a unique solution $Y(t)\in C(\R,E)$.\\
(ii) For every $t\in \R$, the operator $U(t):Y_0\mapsto Y(t)$
is continuous on $E$.\\
(iii) The energy is conserved and finite,
\be\la{3.0}
H(Y(t))=H(Y_0)\,\,\,\mbox{for }\,\,t\in\R.
\ee
\end{lemma}
{\bf Proof.}
{\it Step (i)} In the case when $\rho=0$ the existence and uniqueness
of solution $Y(t)\in C(\R,E)$ to the problem (\ref{1.1'})
is proved by Fourier transform.
Therefore the problem (\ref{1.1'}) for $Y(t)\in C(\R,E)$ 
is equivalent to
\be\la{intD}
Y(t)=e^{{\cal A}_0 t} Y_0+\int\limits_0^t
e^{{\cal A}_0 (t-s)} BY(s)\,ds,
\ee
where 
\beqn
\left.\ba{cc} {\cal A}_0=\left(
\ba{cc}0&I\\ -{\cal H}_0&0
\ea\right), & {\cal H}_0Y^0
=((-\De+m_1^2) \varphi_1,\dots,(-\De+m_d^2) \varphi_d,\omega^2 q),\\
 B(Y^0,Y^1)=(0,RY^0),&
RY^0:=-\Big(q\cdot\nabla\rho_1,\dots,q\cdot\nabla\rho_d,
\ds\sum\limits_{n=1}^d\int \varphi_n(x)\nabla\rho_n(x)\,dx\Big)
\ea\right|\la{A_0}
\eeqn
for $Y^0=(\varphi_1,\dots,\varphi_d,q)$.
Note that $\Vert e^{{\cal A}_0t}Y_0\Vert_E\le C\Vert Y_0\Vert_E$; 
and the second term in (\ref{intD}) is estimated by
$$
\sup\limits_{|t|\le T}\Vert
\int\limits_0^t e^{{\cal A}_0 (t-s)} BY(s)\,ds\Vert_E
\le C\,T\sup\limits_{|s|\le T}\Vert Y(s)\Vert_E. 
$$
This bound and the contraction mapping principle imply
the existence and uniqueness of the local
solution $Y(t)\in C([-\ve,\ve],E)$ with an $\ve>0$.

{\it Step (ii)}
To prove (\ref{3.0}) let us assume that 
$\phi_n^0=(\varphi^0_n,\pi_n^0)\in C^3(\R^3)\times C^2(\R^3)$
and $\phi_n^0(x)=0$ for $|x|\ge R_0$, $n\in\bar d$.
Then
$\varphi_n(x,t)\in C^2(\R^3\times\R)$ and
$$
\varphi_n(x,t)=0\quad\mbox{for }\, |x|\ge |t|+\max\{R_0,R_\rho\}
$$
by the integral representation (\ref{intD})
and condition {\bf A2}. Therefore, for such initial data, the equality
(\ref{3.0}) can be proved by integration by parts. Hence the 
energy conservation (\ref{3.0}) follows from the continuity of $U(t)$
and the fact that 
$[C_0^3(\R^3)]^{d}\oplus \R^3\oplus[C_0^2(\R^3)]^{d}\oplus \R^3$
is dense in $E$.

{\it Step (iii)} Note that 
\beqn
\frac12\Vert\nabla\varphi_n(x)\Vert^2+
\frac12m_n^2\Vert\varphi_n(x)\Vert^2
+\langle\varphi_n(x),q\cdot\nabla\rho_n(x)\rangle\nonumber\\
=\frac1{2(2\pi)^3}
\Big(\Big\Vert \sqrt{k^2+m_n^2}\hat\varphi_n(k)+\frac{1}{\sqrt{k^2+m_n^2}}
iq\cdot k\hat\rho_n(k)\Big\Vert^2
-\Big\Vert\frac{q\cdot k}{\sqrt{k^2+m_n^2}}\hat\rho_n(k)\Big\Vert^2\Big).
\nonumber
\eeqn
Hence the Hamiltonian functional is nonnegative, since
\beqn\la{positive}
H(Y)&=&\frac12\sum\limits_{n=1}^d\Big\{
\Vert\pi_n\Vert^2+\frac1{(2\pi)^3}
\Vert \sqrt{k^2+m_n^2}\hat\varphi_n(k)+\frac{1}{\sqrt{k^2+m_n^2}}
iq\cdot k\hat\rho_n(k)\Vert^2\Big\}\nonumber\\
&&+\frac12\Big(\omega^2|q|^2 -q\cdot K_0q\Big)+\frac12|p|^2\ge0
\eeqn
 by condition {\bf A1'}.
Moreover, by (\ref{3.0}) and (\ref{positive}), for $|t|<\ve$, we obtain that
\beqn\nonumber
\Vert Y(t)\Vert^2_{E}\le C\, H(Y(t))=C\, H(Y_0). 
\eeqn
On the other hand, for $Y_0=(\varphi^0,\pi^0,q^0,p^0)$,
\beqn
H(Y_0)\le\sum\limits_{n=1}^d\left(\Vert\nabla\varphi^0_n\Vert^2+
\frac12m_n^2\Vert\varphi^0_n\Vert^2+
\frac12 \Vert\pi^0_n\Vert^2\right)+
\frac12(\omega^2+\Vert \rho\Vert^2)|q^0|^2+\frac12|p^0|^2.
\eeqn
Hence, we obtain the {\it a priori} estimate
\beqn\la{enest}
\Vert Y(t)\Vert_{E}\le C_1\Vert Y_0\Vert_E\quad\mbox{for }\,\, t\in\R. 
\eeqn
Therefore, properties (i)--(iii) of Lemma \ref{l3.1}
for arbitrary $t\in\R$ follow from the bound (\ref{enest}).
\bo
\medskip

We return to the proof of Proposition \ref{p1.1}.
Let us choose $R>R_\rho$ with $R_\rho$ from condition {\bf A2}.
Then by the integral representation (\ref{intD})
the solution $Y(t)$ for $|x|<R$ depends only on the
initial data $Y_0(x)$ with $|x|<R+|t|$.
Thus the continuity of $U(t)$ in ${\cal E}$ follows 
from the continuity in $E$.  

For every $R>0$ we define  the local energy seminorms by
\beqn\la{3.4'}
\Vert Y\Vert^2_{E(R)}:=
\sum\limits_{n=1}^d\int\limits_{|x|<R}
\Big(|\nabla \varphi_n(x)|^2+m_n^2|\varphi_n(x)|^2+|\pi_n(x)|^2\Big)\,dx
+|q|^2+|p|^2
\eeqn
for $Y=(\varphi,\pi,q,p)$. By the estimate (\ref{enest}), we obtain the 
following local energy estimates:
\be\la{3.2.2}
\Vert U(t) Y_0\Vert^2_{E(R)}\le C\Vert Y_0\Vert^2_{E(R+|t|)},
\ee
for $R>R_\rho$ and $t\in\R$.
Hence, the bound (\ref{3.2.1}) follows from (\ref{3.2.2}).
\bo

\setcounter{equation}{0}
\section{Decay of local energy}
\begin{pro}\la{l5.1}
Let conditions {\bf A1}--{\bf A3} hold and 
let $Y_0\in E$ be such that 
\be\la{**}
\varphi^0(x)=\pi^0(x)=0\,\,\,\mbox{for }\,|x|>R_1,
\ee
with some $R_1>0$. Then
there exists a constant $C=C(R,R_1)>0$ such that
the following bound holds for every $R>0$, 
\be\la{delocen}
\Vert Y(t)\Vert_{{\cal E},R}
\le C\ve_m(t)\Vert Y_0\Vert_{{\cal E},R_1},\quad t\ge0,
\ee
where 
\be\la{ve(t)}
\ve_m(t)=\left\{\ba{ll}
e^{-\delta |t|}\quad \mbox{with a }\, \delta>0,&\mbox{if }\,m=0\\
(1+|t|)^{-3/2}, & \mbox{if }\, m\not=0.
\ea\right.
\ee
\end{pro}

In the case when $m=(m_1,\dots,m_d)=0$ Proposition \ref{l5.1}
is extension of Proposition 7.1 from \cite{KSK},
where a similar result is established in the case when
$d=1$ and $\rho(x)=\rho_r(|x|)$.
In the case when $m\not=0$ we apply the methods of \cite{IKV}.

To prove Proposition \ref{l5.1} we solve the Cauchy problem (\ref{1.1'}) 
applying the Fourier - Laplace transform, 
$$
\ti Y(\lambda)=\ds\int\limits_0^{+\infty}
 e^{-\lambda t} Y(t) \, dt,\,\,\,\,\,\Re \lambda>0.
$$
Then the system (\ref{1})--(\ref{5}) becomes
\beqn
-\varphi_{n}^0(x)+
\lambda\ti\varphi_{n}(x,\lambda)&=&
\ti\pi_{n}(x,\lambda),\quad n=1,\dots,d,\la{1'}\\
\la{2'}
-\pi_{n}^0(x)+ \lambda\ti\pi_{n}(x,\lambda)&=&
(\Delta-m_n^2)\ti\varphi_{n}(x,\lambda)-
\ti q(\lambda)\cdot \nabla\rho_n(x),\\
\la{3'}
-q^0+\lambda \ti q(\lambda)&=&\ti p(\lambda),\\
\la{4'}  
-p^0+\lambda\ti p(\lambda)&=&
-\omega^2\ti q(\lambda)- \sum\limits_{n=1}^d
\int \ti\varphi_n(y,\lambda) \nabla \rho_n(y)\,dy.
\eeqn
From (\ref{1'})--(\ref{4'}) we obtain
\beqn\la{4.13'}
\ti\varphi_n(x,\lambda)
&=&(-\Delta+m_n^2+\lambda^2)^{-1}\left(\lambda\varphi^0_n+\pi^0_n(x)\right)
-(-\Delta+m_n^2+\lambda^2)^{-1} 
\nabla\rho_n(x)\cdot \ti q(\lambda),\nonumber\\
(\lambda^2+\omega^2)\ti q(\lambda)&=&
H(\lambda)\ti q(\lambda)+{\cal R}(\lambda,Y_0),
\eeqn
where
\beqn\la{F}
 {\cal R}(\lambda,Y_0):=
-\sum\limits_{n=1}^d\int (-\Delta+m_n^2+\lambda^2)^{-1}
\left(\pi_n^0(y)+\lambda\varphi^0_n(y)\right) \nabla \rho_n(y)\,dy+
\left(p^0+\lambda  q^0\right),
\eeqn
and $H(\lambda)$ is the $3\times 3$ matrix with the matrix elements
$H_{ij}(\lambda)$, 
\beqn\la{4.9'}
H_{ij}(\lambda)&=&\sum\limits_{n=1}^d
\int\nabla_i \rho_n(y)(-\Delta+m_n^2+\lambda^2)^{-1}\nabla_j\rho_n(y)\,dy.
\eeqn
Therefore,  equation (\ref{4.13'}) rewrites as  
\beqn\la{2.12}
\ti q(\lambda)=\left[(\lambda^2+\omega^2)I-H(\lambda)
\right]^{-1} {\cal R}(\lambda,Y_0)\equiv
\ti{\cal N}(\lambda){\cal R}(\lambda,Y_0),
\eeqn
where by $\ti{\cal N}(\lambda)$
we denote a $3\times 3$ matrix  of the form
\be\la{A}
\ti{\cal N}(\lambda)=D^{-1}(\lambda),\,\,\,\,D(\lambda):=
(\lambda^2+\omega^2)I-H(\lambda)
\,\,\,\,\mbox{for }\,\, \Re\lambda>0.
\ee

\subsection{Time decay for $q(t)$ and $p(t)$}
In this subsection we prove the exponential decay
for $q(t)$ and $p(t)$ in the case when  $m=0$.
The case $m\not=0$ is considered in Appendix.
\begin{theorem}\la{thm=0}
Let $m=0$, conditions {\bf A1}--{\bf A3} and (\ref{**}) hold.
 Then there exists a $\delta>0$ such that
the following bound holds 
\beqn\la{p+q}
|q(t)|+|p(t)|\le Ce^{-\delta t}\Vert Y_0\Vert_{{\cal E},R_1}.
\eeqn 
\end{theorem}

To prove Theorem \ref{thm=0} we first investigate
the properties of the matrix $D(\lambda)$.

Denote $\C_\beta:=\{\lambda\in \C:\,\Re\lambda>\beta\}$ for $\beta\in\R$.
\begin{lemma}\la{detA}
Let $m=0$ and conditions {\bf A1}--{\bf A3} hold. Then
(i) $D(\lambda)$ admits an analytic continuation to $\C$;
(ii) for every $\beta>0$ $\exists N_\beta>0$
such that $v\cdot D(\lambda)v\ge C|v|^2|\lambda|^2$ 
for $\lambda\in \C_{-\beta}$ 
with $|\lambda|\ge N_\beta$ and every $v\in\R^3$.
(iii) There exists a $\delta>0$ such that 
$v\cdot D(\lambda)v\not=0$ for 
 $\lambda\in \overline\C_{-\delta}$ and for every $v\not=0$.
\end{lemma}

Lemma \ref{detA}  is a modification of Lemma 7.2 of
 \cite{KSK}, where the result is proved 
in the case when $d=1$ and $\rho(x)=\rho_r(|x|)$.\\
{\bf Proof}.
(i) We rewrite the entries of the matrix $H(\lambda)$ as
\be\la{Mij}
H_{ij}(\lambda)=
\sum\limits_{n=1}^d\int\int \nabla_i \rho_n(y)
\frac{e^{-\lambda|y-z|}}{4\pi|y-z|}
\nabla_j \rho_n(z)\,dydz.
\ee
Hence, $H_{ij}(\lambda)$ is defined and has an analytic continuation to $\C$.
Therefore property (i) follows.

(ii) From (\ref{Mij}) it follows that 
\be\la{4.11}
H_{ij}(\lambda)\to0 \,\,\,\,\,\mbox{as }\,|\lambda|\to\infty
\,\,\,\,\mbox{with }\,\,\lambda\in\C_{-\beta}
\ee
what implies (ii).

(iii) At first note that the matrix $D(\lambda)$ is positive definite
if $\Im\lambda=0$.
Indeed, from  condition {\bf A1}
it follows that, for any $v\in \R^3\setminus\{0\}$,
\beqn
v\cdot D(\lambda)v&=& 
(\lambda^2+\omega^2)|v|^2-\sum\limits_{n=1}^d\frac1{(2\pi)^3}\int
\frac{(v\cdot k)^2}{k^2+\lambda^2}
|\hat\rho_n(k)|^2\,dk\nonumber\\
&\ge&\om^2|v|^2-\sum\limits_{n=1}^d\frac1{(2\pi)^3}
\int\frac{(v\cdot k)^2}{k^2}|\hat\rho_n(k)|^2\,dk>0.\nonumber
\eeqn
Secondly, for $y\in\R^1\setminus\{0\}$, we find
\beqn
H_{ij}(iy+0)=\sum\limits_{n=1}^d\frac1{(2\pi)^3}\int\limits_{\R^3}
\frac{k_ik_j|\hat\rho_n(k)|^2}{k^2-(y-i 0)^2}dk
=\int\limits_{0}^{+\infty}
\frac{r^4g_{ij}(r)}{(r-y+i 0)(r+y-i 0)}\,dr,\nonumber 
\eeqn
where
\beqn\nonumber
g_{ij}(r):=\sum\limits_{n=1}^d
\frac1{(2\pi)^3}\int\limits_{|\theta|=1}
\theta_i\theta_j |\hat\rho_n(r\theta)|^2\,dS_\theta.
\eeqn
Note that
$g_{ij}\in C^\infty([0,+\infty))$,
$\ds\max_{r\in[0,\infty)}(1+r)^{N}|g_{ij}(r)|<\infty$ for every $N>0$
by condition {\bf A2}.
Applying the Plemelj formula (see \cite{GSh}) and
condition {\bf A3} we obtain
\beqn\la{4.14'}
v\cdot\Im H(iy+0) v=-\pi \frac{y^3}{2} 
\sum\limits_{n=1}^d
\frac1{(2\pi)^3}\int\limits_{|\theta|=1}
(v\cdot\theta)^2|\hat\rho_n(|y|\theta)|^2\,dS_\theta\not =0
\eeqn
for any vector $v\in\R^3\setminus\{0\}$. Hence
$v\cdot\Im D(iy+0)v=- v\cdot\Im H(iy+0) v\not=0.$
Lemma \ref{detA} is proved. \bo
\medskip

Denote by ${\cal N}(t)$ an inverse Laplace transformation of 
$\ti{\cal N}(\lambda)$,
\be\la{4.15}
{\cal N}(t)=\frac{1}{2\pi i}\int\limits_{-i\infty-\delta}^{i\infty-\delta}
e^{\lambda t}\ti{\cal N}(\lambda)\,d\lambda\,\,\,\,\,\mbox{for }\,t>0.
\ee
\begin{lemma}\la{c1}
Let $m=0$ and conditions {\bf A1}--{\bf A3} hold. Then for $j=0,1,\dots$, 
\be\la{decayA}
|{\cal N}^{(j)}(t)|\le C e^{-\delta t},\,\,\,\,\,\,t>1.
\ee
\end{lemma}
{\bf Proof}.
By Lemma \ref{detA} the bound on ${\cal N}(t)$ follows.
To prove the bound for $\dot{\cal N}(t)$ we consider 
$\lambda\ti{\cal N}(\lambda)$ and prove the following bound:
\be\la{4.14}
\left|v\cdot (\lambda\ti{\cal N}(\lambda))' v\right|
\le\frac{C\, |v|^2}{1+|\lambda|^{2}},\quad
\mbox{for }\, \lambda\in\overline\C_{-\delta}.
\ee
This implies that
$
\max\limits_{t\in[0,+\infty)}|t e^{\delta t}\dot{\cal N}(t)|<\infty,
$
and the bound (\ref{decayA}) for $\dot{\cal N}(t)$ follows.
To prove (\ref{4.14}) it suffices to establish that
$|\ti{\cal N}'_{ij}(\lambda)|\le C(1+|\lambda|)^{-3}$.
From (\ref{Mij}) it follows  that
$$
H'_{ij}(\lambda)\to0 \,\,\,\,\,\mbox{as }\,|\lambda|\to\infty
\,\,\,\,\mbox{with }\,\,\lambda\in\C_{-\delta}.
$$
Therefore, Lemma \ref{detA} and (\ref{A}) imply that, 
for $i,j=1,2,3$,
$$
|\ti{\cal N}'_{ij}(\lambda)|\le \frac{C}{1+|\lambda|^4}
\Big(\sum\limits_{k,l=1}^3 |D'_{kl}(\lambda)|\Big)
\le \frac{C_1}{1+|\lambda|^{3}}\quad \mbox{as }\,|\lambda|\to\infty.
$$
This implies (\ref{4.14}).
The bound (\ref{decayA}) with $j\ge2$ is proved similarly. \bo
\medskip\\
{\bf Proof of Theorem \ref{thm=0}}.
Denote by $\varphi_n^1(t)\equiv\varphi_n^1(x,t)$ 
the solution of the Cauchy problem (\ref{freeKG})
with the initial data $\phi_n^0=(\varphi_n^0,\pi_n^0)$.
Then (\ref{F}) and (\ref{2.12}) imply that for $t>0$
\beqn \la{5.10}
q(t)=I(t,\phi^0)+R(t,q^0,p^0),
\eeqn
where $\phi^0=(\phi_1^0,\dots,\phi^0_d)$ and
\beqn
I(t,\phi^0)&:=&
-\sum\limits_{n=1}^{d} \int\limits_0^t
\langle \varphi_n^1(t-s), {\cal N}(s)\nabla\rho_n\rangle\,ds
\nonumber\\
&=&\sum\limits_{n=1}^{d} \int\limits_0^t
\langle {\cal N}(s) \nabla \varphi_n^1(t-s),\rho_n\rangle\,ds,
\la{Phi1}\\
R(t,q^0,p^0)&:=&{\cal N}(t)p^0+\dot {\cal N}(t)q^0.  \la{Phi2}
 \eeqn
First, by the bound (\ref{decayA})  we have
\be\la{decayQ}
|R(t,q^0,p^0)|\le C e^{-\delta t}(|q^0|+|p^0|).
\ee
Secondly, since $m=0$, 
\be\la{freewave}
W_n(t)\phi_n^0=0\,\,\, \mbox{ for }\, x\in B_R\quad \mbox{and }\,\,t>R+R_1.
\ee
due to the condition (\ref{**}) and the strong Huygens principle. 
Hence, from (\ref{decayA}) and (\ref{Phi1}) we obtain (see (\ref{enest}))
that, for $t>R_1+R_\rho$,
\beqn
|I(t,\phi^0)|&\le& \sum\limits_{n=1}^d
C(\rho_n)\int\limits_0^t |{\cal N}(s)|
\Vert \nabla \varphi_n^1(t-s)\Vert_{L^2(B_{R_\rho})}\,ds
\nonumber\\
&\le&\sum\limits_{n=1}^d C_1\int\limits_{t-R_1-R_\rho}^t 
e^{-\delta s}\Vert\phi_n^0\Vert_{R_1}\,ds
\le C e^{-\delta t} \Vert\phi^0\Vert_{R_1}.\la{5.20}
\eeqn
Relations (\ref{5.10}), (\ref{decayQ}) and (\ref{5.20})
imply the bound (\ref{p+q}) for $q(t)$. The bound for $p(t)$
is proved similarly. \bo

\subsection{Time decay of field components}
Now we finish the proof of Proposition \ref{l5.1}.
From equations (\ref{1}) we have
\beqn\la{6.9.1}
\phi_n(x,t)&=&W_n(t)\phi^0_n-\int\limits_0^tW_n(t-s)\left(
\ba{cc}0\\\nabla\rho_n(x)\ea\right)\cdot q(s)\,ds,\,\,\,n\in\bar d.
\eeqn
Let $m_n=0$. Then the bound (\ref{freewave}),
the condition (\ref{**}) and Theorem \ref{thm=0} imply
\beqn\nonumber
\Vert\phi_n(t)\Vert_R\le C\int\limits_{t-R_{\rho}-R}^t
e^{-\delta s}\Vert Y_0\Vert_{{\cal E},R_1}\,ds
\le Ce^{-\delta t} \Vert Y_0\Vert_{{\cal E},R_1}
\eeqn
for $t>R+\max\{R_\rho,R_1\}$.
Let $m_n\not=0$. Then instead of (\ref{freewave}) we apply the following 
well-known bound:
$$
\Vert W_n(t)\phi^0_n\Vert_R
\le C (1+t)^{-3/2}\Vert\phi_n^0\Vert_{R_1},\quad t\ge0.
$$
Hence  Theorem \ref{Appendix} and the representation
(\ref{6.9.1}) yield
\beqn
\Vert\phi_n(t)\Vert_R
\le  C(1+t)^{-3/2}\Vert Y_0\Vert_{{\cal E},R_1}.\nonumber
\eeqn
The bound (\ref{delocen}) is proved. \bo

\setcounter{equation}{0}
\section{Compactness of measures $\mu_t$}
\begin{lemma}
Let conditions {\bf A1}--{\bf A3} and {\bf S0}--{\bf S2} hold. Then
\be\la{7.1.1}
\sup\limits_{t\ge 0} E\Vert U(t) Y_0\Vert^2_{{\cal E},R}
\le C(R)<\infty,\,\,\,\forall R>0.
\ee
\end{lemma}
{\bf Proof}
Let us write $U_0(t)=e^{{\cal A}_0 t}$ (see (\ref{A_0})). 
At first, note that
\be\la{7.1.10}
\sup\limits_{t\ge 0} E\Vert U_0(t) Y_0\Vert^2_{{\cal E},R}
\le C(R),\,\,\,\forall R>0.
\ee
Indeed (see (\ref{1.5}) and (\ref{2.1'})),
$$
\Vert U_0(t) Y_0\Vert^2_{{\cal E},R}=
\Vert W(t) \phi^0\Vert^2_{R}+|q_0(t)|^2+|\dot q_0(t)|^2,
$$
where the operator $W(t)$ is defined in (\ref{W(t)})
and $q_0(t)$ is a solution to the Cauchy problem
$$
\ddot q_0(t)+\omega^2 q_0(t)=0,\,\,\,t\in\R,\,\,\,\,
(q_0(t),\dot q_0(t))|_{t=0}=(q^0,p^0).
$$
Hence, $|q_0(t)|+|\dot q_0(t)|\le C(|q^0|+|p^0|)$.
From \cite[Proposition 3.2]{DKKS} and \cite[Proposition 3.1]{DKRS} 
it follows that
\be\la{6.3}
\sup\limits_{t\ge 0} E\Vert W(t) \phi^0\Vert^2_{R}=
\sup\limits_{t\ge 0} \sum\limits_{n=1}^d
E\Vert W_n(t) \phi_n^0\Vert^2_{R}\le C(R),
\,\,\,\forall R>0.
\ee
It implies the bound (\ref{7.1.10}).
Further, we represent the solution to (\ref{1.1'}) as follows
\be\la{6.4}
U(t)Y_0=U_0(t) Y_0+\int\limits_0^t
U(t-s) BU_0(s)Y_0\,ds,
\ee
where the operator $B$ is defined in (\ref{A_0}).
Hence, (\ref{delocen}) and (\ref{7.1.10}) yield
\beqn
E\Vert U(t)Y_0\Vert^2_{{\cal E},R}&\le& E\Vert U_0(t) Y_0\Vert^2_{{\cal E},R}
+E\int\limits_0^t \Vert U(t-s) BU_0(s)Y_0\Vert^2_{{\cal E},R}\,ds
\nonumber\\
&\le& C(R)+\int\limits_0^t \ve^2_m(t-s)
E\Vert U_0(s) Y_0\Vert^2_{{\cal E},R_\rho}\,ds\le C_1(R)<\infty.\,\,\,\bo
\nonumber
\eeqn

\setcounter{equation}{0}
\section{Convergence of characteristic functionals}

\subsection{Asymptotic behavior of $Y(t)$}
\begin{pro}\la{l7.1}
Let conditions {\bf A1}--{\bf A3} and {\bf S0}--{\bf S3} hold. Then 

(i) the following bounds hold,
\beqn\la{8.1}
E|q_i(t)-\sum\limits_{k=1}^d
\langle W_k(t)\phi_k^0,\alpha_k^i\rangle|^2
&\le&C\ti\ve_m(t),\\
 E|p_i(t)-\sum\limits_{k=1}^d
\langle W_k(t)\phi_k^0,\beta_k^i
\rangle|^2&\le & C\ti\ve_m(t),\,\,\,\,t>1,  \la{8.2}
\eeqn
where $\ti\ve_m(t)= e^{-2\delta t}$, if $m=0$, and
$\ti\ve_m(t)= (1+t)^{-1}$ otherwise,
the functions $\alpha_k^i,\beta_k^i$ 
 are defined in (\ref{al}) and (\ref{beta}), $i=1,2,3$.

(ii) Let $\psi\in [C_0^\infty(\R^3)]^2$ 
with $\supp \psi\subset B_R$.  Then, for $n\in \bar d$
and $t\ge1$,
\be\la{8.3}
E\Big|\langle\phi_n(t),\psi\rangle
- \left\langle W_n(t)\phi^0_n,\psi\right\rangle+
\sum\limits_{k=1}^d
\left\langle W_k(t)\phi_k^0,\theta_{kn}\right\rangle\Big|^2
\le C\ti\ve_m(t),
\ee
where the functions $\theta_{kn}(x)$, $k,n\in \bar d$, 
are defined in (\ref{teta}) with $\psi_n=\psi$.
\end{pro}
{\bf Proof.} (i) 
At first, relations (\ref{5.10})--(\ref{Phi2}),
the bounds (\ref{decayQ}) and (\ref{7.10}) yield
\be\la{8.4}
E \Big|q_i(t)+\sum\limits_{k=1}^d\sum\limits_{r=1}^3\int\limits_0^t
\left\langle \varphi^1_k(t-s),{\cal N}_{ir}(s)\nabla_r\rho_k
\right\rangle\,ds\Big|^2 \le C\ve^2_m(t)
\ee
with $\ve_m(t)$ from (\ref{ve(t)}). Secondly,
\beqn
&& E\left|\int\limits_t^{+\infty}
\left\langle \varphi^1_k(t-s),{\cal N}_{ir}(s)\nabla_r\rho_k
\right\rangle\,ds\right|^2\nonumber\\
&=&\!\! \int\limits_t^{+\infty}\! {\cal N}_{ir}(s_1)\,ds_1
\int\limits_t^{+\infty}\!{\cal N}_{ir}(s_2)
E\left(
\langle\varphi^1_k(t-s_1),\nabla_r\rho_k\rangle
\langle \varphi^1_k(t-s_2),\nabla_r\rho_k\rangle
\right)\,ds_2.\,\,\,\,\nonumber
\eeqn
For any $t,s_1, s_2\in\R$, we have
\beqn
\left|E\left(
\langle \varphi^1_k(t-s_1),\nabla_r\rho_k\rangle 
\langle\varphi^1_k(t-s_2),\nabla_r\rho_k\rangle
\right)\right|&\le& C
\sup_{\tau\in\R}E|\langle \varphi^1_k(\tau),\nabla_r\rho_k\rangle|^2
\nonumber\\
&\le& C_1\sup_{\tau\in\R}E\Vert \varphi^1_k(\tau)\Vert^2_{L^2(B_{R_\rho})}
\le C_2<\infty\nonumber
\eeqn
by the bound (\ref{6.3}).
Hence, applying the bounds (\ref{decayA}) and (\ref{7.10}) we obtain  that
\be\la{8.7}
E\left|\int\limits_t^{+\infty} 
\langle\varphi^1_k(t-s),{\cal N}_{ir}(s)\nabla_r\rho_k\rangle
\,ds\right|^2\le C\ti\ve_m(t).
\ee
Therefore, (\ref{8.1}) follows from (\ref{8.4}), (\ref{8.7}) and (\ref{al}), 
since 
\beqn\nonumber
\left\langle \varphi^1_k(t-s),{\cal N}_{ir}(s)\nabla_r\rho_k\right\rangle
= \left\langle W_k(t)\phi_k^0,
W'_k(-s)\left(\ba{c} {\cal N}_{ir}(s)\nabla_r\rho_k\\0\ea
\right) \right\rangle.
\eeqn
The bound (\ref{8.2}) can be proved by similar way.
\medskip

(ii) Let $\psi\in [C_0^\infty(\R^3)]^2$ with $\supp \psi\subset B_R$. 
From (\ref{6.9.1}) it follows that 
\beqn\la{6.7}
\langle \phi_n(t),\psi\rangle=
\langle W_n(t)\phi_n^0,\psi\rangle-
\sum\limits_{i=1}^3 \int\limits_0^{t}
q_i(t-s)\left\langle W_n(s) {\cal R}_{in},\psi \right\rangle\,ds,\,\,\,\,
{\cal R}_{in}\equiv\left(\ba{c}0\\\nabla_i\rho_n\ea\right).
\eeqn
Note that
\beqn\la{6.8}
\langle W_n(s) {\cal R}_{in},\psi\rangle=\left\{
\ba{ll}0\,\,\, \mbox{for }\,s>R_\rho+R&\mbox{if }\, m_n=0,\\
{\cal O}((1+s)^{-3/2})&\mbox{if }\, m_n\not=0.
\ea\right.
\eeqn
Then from (\ref{8.1}) and (\ref{6.8}) it follows that
\beqn\la{7.8}
E\left|\int\limits_0^{t}
\Big(q_i(t-s)-\sum\limits_{k=1}^d
\langle W_k(t-s)\phi_k^0,\alpha^i_k\rangle\Big)
\left\langle W_n(s){\cal R}_{in},\psi\right\rangle
\,ds\right|^2\le C\ti\ve_m(t).
\eeqn
Denote
\beqn\nonumber
I_n(t)&:=&E\Big|\int\limits_t^\infty\sum\limits_{k=1}^d
\langle W_k(t-s)\phi_k^0,\alpha_k^i\rangle
\langle W_n(s){\cal R}_{in},\psi\rangle\,ds\Big|^2.
\eeqn
If $m_n=0$, $I_n(t)=0$ for $t>R_\rho+R$ by (\ref{6.8}).
If $m_n\not=0$, 
\be\la{6.9}
|I_n(t)|\le C\ti\ve_m(t).
\ee
It follows from the following estimate: for $\tau\in\R$,
\beqn\nonumber
E|\langle W_k(\tau)\phi_k^0,\alpha_k^i\rangle|^2
=E\Big|\sum\limits_{r=1}^d\int\limits_0^{+\infty}
{\cal N}_{ir}(s)\left\langle W_k(\tau-s)\phi_k^0,
\left(\ba{c}\nabla_r\rho_k\\0\ea
\right)\right\rangle\,ds\Big|^2 \le C<\infty,
\eeqn
by (\ref{6.3}) and (\ref{decayA}) and (\ref{7.10}). 
The relation ({\ref{6.7}) and the bounds (\ref{7.8}) and (\ref{6.9})
imply the bound (\ref{8.3}). \bo
\begin{cor}\la{c7.2}
Let $Z=(\psi,u,v)\in {\cal D}={\cal D}_0\times\R^3\times\R^3$.
Then
\beqn\nonumber
\langle Y(t),Z\rangle=
 \langle W(t) \phi^0, \psi^Z\rangle + r(t),
\eeqn
where $\langle Y(t),Z\rangle= \langle \phi(t),\psi\rangle 
+ q(t)\cdot u+p(t)\cdot v$, 
$Y(t)=(\phi(t),q(t),p(t))$ is a solution to the Cauchy problem (\ref{1.1'}), 
the function $\psi^Z$ is defined in (\ref{hn})
 and $E|r(t)|^2\le C \ti\ve_m(t)$.
\end{cor}

\subsection{The end of the proof of Theorem 2.11}
\begin{pro}
Let all assumptions of Theorem \ref{tA} be fulfilled. Then,
for $Z\in{\cal D}$,
$$
E \exp\{i\langle Y(t),Z\rangle\}\to
\exp\left\{-\frac{1}{2} {\cal Q}_\infty(Z,Z)\right\},\,\,\,t\to\infty.
$$
\end{pro}
{\bf Proof}. By triangle inequality we have
\beqn\la{8.16}
&&\left|E e^{i\langle Y(t),Z\rangle}-
\exp\{-\frac{1}{2} {\cal Q}_\infty(Z,Z)\}\right|
\le\Big|E \Big(e^{i\langle Y(t),Z\rangle}-
 e^{i\langle W(t)\phi^0,\psi^Z\rangle}\Big)\Big|
\nonumber\\
&&+ \Big|E e^{i\langle W(t)\phi^0,\psi^Z\rangle}
-\exp\{-\frac{1}{2}{\cal Q}_\infty (Z,Z)\}\Big|.
\eeqn
The first term in the RHS of (\ref{8.16}) is estimated by
\beqn
&&\Big|E \Big(e^{i\langle Y(t),Z\rangle}-
e^{i\langle W(t)\phi^0,\psi^Z\rangle}\Big)\Big|
\le E\Big|\langle Y(t),Z\rangle
-\langle W(t)\phi^0,\psi^Z\rangle\Big|
\nonumber\\
&&\le E|r(t)|
\le \Big(E|r(t)|^2\Big)^{1/2} \le C \ti\ve^{1/2}_m(t)\to0\,\,\,\,\,
\mbox{as }\,t\to\infty,
\eeqn
by Corollary \ref{c7.2}. It remains to prove the convergence of
$E\exp\{i\langle W(t)\phi^0,\psi^Z\rangle\}\equiv\hat\nu_t(\psi^Z)$
to a limit as $t\to\infty$.
In \cite{DKKS, DKRS} we proved the convergence of $\hat\nu_t(\psi)$
to a limit for $\psi\in{\cal D}_0$. 
However, generally, $\psi^Z\not\in{\cal D}_0$.
Now we introduce a space $H_m$ such that $\psi^Z\in H_m$
and the characteristic functionals $\hat\nu_t(\psi)$, $t\in\R$,
are equicontinuous in $H_m$.

Denote $\omega_n(k)=(|k|^2+m_n^2)^{1/2}$, $n\in\bar d$.
\begin{definition}
 $H_m$ is the space of the pairs $\psi=(\psi^0,\psi^1)$
of $\R^d$-valued functions $\psi^0=(\psi^0_1,\dots,\psi^0_d)$ and 
$\psi^1=(\psi^1_1,\dots,\psi^1_d)$, such that
$\psi^0_n$, $\omega_n^{-1}\hat\psi_n^0\in L^2(\R^3)$,
$\psi_n^1\in H^1(\R^3)$, $n\in\bar d$,  
 with the finite norm   
$$
\Vert\psi\Vert^2_m:=\sum\limits_{n=1}^d\left(\Vert \psi_n^0\Vert^2
+\Vert\omega_n^{-1}(k)\hat \psi_n^0(k)\Vert^2
+\Vert\psi_n^1\Vert^2+\Vert\nabla\psi_n^1\Vert^2\right).
$$
\end{definition}

The formulas (\ref{hn})--(\ref{teta}) and condition {\bf A2}
imply that the functions $\psi^Z$ satisfying the bound
$\sup_{k\in\R^3}(1+|k|)^N|\hat\psi^Z(k)|<\infty$ for every $N>0$,
and $\psi^Z\in H_m$. Note that if all $m_n\not=0$,
$H_m=L^2(\R^3)\oplus H^1(\R^3)$.
\begin{lemma}\la{l}
(i) The quadratic form ${\cal Q}^\nu_t(\psi,\psi)=\ds\int|\langle
\phi^0,\psi\rangle|^2\nu_t(d\phi^0)$, $t\in\R$,
are equicontinuous in $H_m$.\\
(ii) The characteristic functionals $\hat\nu_t(\psi)$, $t\in\R$,
are equicontinuous in $H_m$.
\end{lemma}
{\bf Proof}. (i) It suffices to prove the uniform bounds
\be\la{6.12}
\sup\limits_{t\in\R} |{\cal Q}^\nu_t(\psi,\psi)|\le C
\Vert\psi\Vert_m^2,\quad \psi\in H_m.
\ee
Note that
$ {\cal Q}^\nu_t(\psi,\psi)
=\ds\sum\limits_{n,n'=1}^d
\langle q_{0,nn'}(x-y), W'_n(t)\psi_n(x)\otimes W'_{n'}(t)\psi_{n'}(y)\rangle$.
Then by Remark \ref{r2.9} we have
$$
\sup\limits_{t\in\R} |{\cal Q}^\nu_t(\psi,\psi)|
\le C \sup\limits_{t\in\R} \sum\limits_{n=1}^d
\Vert W'_n(t)\psi_n\Vert_{L^2}^2\le C\Vert\psi\Vert_m^2,
$$
since in the Fourier transform we have
$$
F_{x\to k}[ W'_n(t)\psi]=\left(
\ba{ccc} \cos\omega_n(k)t&-\omega_n(k)\sin\omega_n(k)t\\
\omega_n^{-1}(k)\sin\omega_n(k)t&\cos\omega_n(k)t
\ea\right)\left(\ba{c}\hat\psi^0(k)\\\hat\psi^1(k)\ea\right),\,\,\,\,\,
\mbox{for }\,\psi=(\psi^0,\psi^1).
$$
(ii) By the Cauchy-Schwartz inequality and (\ref{6.12}) we obtain that
$$
\ba{rcl}
|\hat\nu_t(\psi_1)-\hat\nu_t(\psi_2)|&=&
|\ds\int \Big( e^{i\langle \phi^0,\psi_1 \rangle}-
e^{i\langle \phi^0,\psi_2 \rangle}\Big)\nu_t(d\phi^0)|
\le
\ds\int |e^{i\langle \phi^0,\psi_1-\psi_2 \rangle}-1|\nu_t(d\phi^0)\\
&\le&
\ds\int |\langle \phi^0,\psi_1-\psi_2 \rangle|\nu_t(d\phi^0)
\le \sqrt {\ds\int |\langle \phi^0,\psi_1-\psi_2 \rangle|^2\nu_t(d\phi^0)}\\
&=&
\sqrt {{\cal Q}^\nu_t(\psi_1-\psi_2, \psi_1-\psi_2)}
\le C\Vert\psi_1-\psi_2 \Vert_m.
~~~~~~~~~~\bo
\ea
$$

Since $\psi^Z\in H_m$,
Proposition 3.3 of \cite{DKKS} (or Proposition 3.2 of \cite{DKRS})
and Lemma \ref{l}, (ii) yield 
$$
\hat\nu_t(\psi^Z)
\to \exp\{-\frac{1}{2}{\cal Q}^\nu_{\infty}(\psi^Z,\psi^Z)\}\,\,\,\,
\mbox{as }\,t\to\infty,
$$
where ${\cal Q}^\nu_{\infty}$ is defined by (\ref{qpp}).
This completes the proof of Theorem \ref{tA}. \bo 

\subsection{Convergence of correlation functions}\la{s.conv}
\begin{pro}
Let all assumptions of Theorem \ref{tA} be fulfilled. Then,
for $Z_1,Z_2\in {\cal D}$,
\be\la{concorf}
E\left(\langle Y(t),Z_1\rangle\langle Y(t),Z_2\rangle\right)
\to {\cal Q}_\infty(Z_1,Z_2),\,\,\,t\to\infty.
\ee
\end{pro}
{\bf Proof}. It is enough to prove the convergence of 
$E|\langle Y(t),Z\rangle|^2$ to a limit as $t\to\infty$.
From Corollary \ref{c7.2} it follows that, for $Z\in {\cal D}$,
\beqn
E|\langle Y(t),Z\rangle|^2=
E|\langle W(t)\phi^0,\psi^Z\rangle|^2+o(1)
={\cal Q}^\nu_t(\psi^Z,\psi^Z)+o(1),
\,\,\,t\to\infty,\nonumber
\eeqn
where $\psi^Z$ is defined in (\ref{hn}) and $\psi^Z\in H_m$.
Therefore, Proposition 6.2 of \cite{DKKS} (or Lemma 4.4 of \cite{DKRS})
and Lemma \ref{l}, (i) imply  that
$\lim_{t\to\infty}{\cal Q}^\nu_t(\psi^Z,\psi^Z)
={\cal Q}^\nu_\infty(\psi^Z,\psi^Z)$.
Formula (\ref{Qmu}) implies (\ref{concorf}). \bo
\medskip

{\bf Acknowledgement} 

T.D. thanks Prof. A. Komech for helpful discussions.

\setcounter{equation}{0}
\section{ Appendix: Time decay of $q(t)$, $p(t)$: 
Case $m\not=0$}
Here we prove the following result, which is a modification 
of Lemma 17.1 of \cite{IKV}.
\begin{theorem}\la{Appendix}
Let $m\not=0$, the conditions {\bf A1}--{\bf A3} hold,
and $Y_0\in E$ with $\phi^0(x)=0$ for $|x|>R_1$. Then $q(t)$
and $p(t)$ are continuous and the following bound holds 
$$
|q(t)|+|p(t)|\le C(1+t)^{-3/2},\quad t\ge0.
$$
\end{theorem}

Ar first, we rewrite the system (\ref{1'})--(\ref{4'}) in the form
\be\la{a1}
({\cal A}-\lambda)\ti Y(\lambda)=-Y_0, \,\,\,\,\Re\lambda>0,
\ee
where ${\cal A}={\cal A}_0+B$ and the operators ${\cal A}_0$, $B$
are defined in (\ref{A_0}).
Hence, the solution $\ti Y$ is given by
$$
\ti Y(\lambda)=-({\cal A}-\lambda)^{-1}Y_0,\,\,\,\,\Re\lambda>0,
$$
if the resolvent $R(\lambda)=({\cal A}-\lambda)^{-1}$ exists
for $\Re\lambda>0$.
\begin{pro} 
The operator-valued function $R(\lambda):E\to E$ is analytic
for $\Re\lambda>0$.
\end{pro}
{\bf Proof}. It suffices to prove that the operator
$({\cal A}-\lambda):E\to E$ has a bounded inverse operator 
for $\Re\lambda>0$.

Let us prove that Ker$({\cal A}-\lambda)=0$ for $\Re\lambda>0$.
Indeed, let $\ti Y(\lambda)\in E$ is a solution (\ref{a1}) 
with $Y_0=0$.
The function $Y(t)=\ti Y(\lambda) e^{\lambda t}\in C(\R,E)$
is the solution to the equation $\dot Y={\cal A}Y$.
Then the Hamiltonian $H(Y(t))=e^{2\lambda t}H(\ti Y(\lambda))$ 
grows exponentially by (\ref{positive}). 
This grow contradicts (\ref{3.0}).
Hence, (\ref{positive}) implies that $\ti\pi(x,\lambda)=0$, $\ti p(\lambda)=0$.
Equations (\ref{1'}) and (\ref{3'}) imply that
$\ti\varphi(x,\lambda)=0$ and $\ti q(\lambda)=0$ because $\lambda\not=0$.

Remember that ${\cal A}={\cal A}_0+B$ (see (\ref{A_0})), where
the operator $B$ is finite-dimensional and the operator ${\cal A}^{-1}_0$
is bounded in $E$.  We rewrite 
$$
{\cal A}-\lambda=({\cal A}_0-\lambda)(I+({\cal A}_0-\lambda)^{-1}B),
$$
where $({\cal A}_0-\lambda)^{-1}B$ is a compact operator.
Since Ker$(I+({\cal A}_0-\lambda)^{-1}B)=0$, the operator
$I+({\cal A}_0-\lambda)^{-1}B$ is invertible by the Fredholm theory.\bo
\medskip

Denote by $g_{\lambda,n}$ 
the fundamental solution of the operator $-\Delta+m_n^2+\lambda^2$,
\be\la{gn}
g_{\lambda,n}(y)=\frac{e^{-\kappa_n|y|}}{4\pi|y|},\,\,\,\,\,n\in \bar d,
\ee
where $\kappa_n^2=m_n^2+\lambda^2$, $\Re\kappa_n>0$ for $\Re\lambda>0$.
From (\ref{1'})--(\ref{4'}) it follows that
$$
M(\lambda)\left(\ba{c}\ti q(\lambda)\\ \ti p(\lambda)\ea\right)=
\left(\ba{c} q^0\\  \bar p^0(\lambda)\ea\right),
$$
where $\bar p^0(\lambda)=p^0-\ds\sum\limits_{n=1}^d
\langle g_{\lambda,n}*(\lambda \ti\varphi^0_n(\lambda)+\ti\pi^0_n(\lambda)),
\nabla\rho_n\rangle$,
\be\la{M(lambda)}
M(\lambda):=\left(\ba{cc}
\lambda I&-I\\ \omega^2I-H(\lambda)&\lambda I\ea\right),
\ee
and the matrix $H(\lambda)$ is defined in (\ref{4.9'}).
The entries of $H(\lambda)$ are of the form  
\beqn\la{M}
H_{ij}(\lambda)= \sum\limits_{n=1}^d
\int\nabla_i \rho_n(y)(g_{\lambda,n}*\nabla_j\rho_n)(y)\,dy
=\sum\limits_{n=1}^d\frac{1}{(2\pi)^3}
\int\frac{k_ik_j|\hat\rho_n(k)|^2}{k^2+m_n^2+\lambda^2}\,dk.
\eeqn
The following result is proved in \cite[p.351]{IKV}.
\begin{lemma}\la{l7.3} 
(i) The operator $-\Delta+m_n^2+\lambda^2$ is invertible in $L^2(\R^3)$
for $\Re\lambda>0$ and its fundamental solution (\ref{gn}) 
decays exponentially as $|y|\to\infty$.

(ii)  For every fixed $y\not=0$, the Green function $g_{\lambda,n}(y)$
admits an analytic continuation (in variable $\lambda$) to the Riemann 
surface of the algebraic function $\sqrt{\lambda^2+m_n^2}$
with the branching points $\pm im_n$ if $m_n\not=0$.
\end{lemma}

Lemma \ref{l7.3} and notation (\ref{M(lambda)}) imply that
 $M(\lambda)$ admits an analytic continuation from the domain
$\Re\lambda>0$ on the Riemann surface with the branching points $\pm im_n$
with $m_n\not=0$, $n\in\bar d$.
 Moreover, 
the matrix $M^{-1}(\lambda)$ exists for large $\lambda$.
It follows from (\ref{M(lambda)}) since $H(\lambda)\to0$
as $\Re\lambda\to\infty$ by (\ref{M}).
\begin{cor}
(i) The matrix $M(\lambda)$ is invertible for $\Re\lambda>0$, and
\be\la{7.5}
\left(\ba{c}\ti q(\lambda)\\ \ti p(\lambda)\ea\right)=
M^{-1}(\lambda)\left(\ba{c} q^0\\  \bar p^0(\lambda)\ea\right),
\,\,\,\,\Re\lambda>0.
\ee
(ii) The matrix $M^{-1}(\lambda)$ admits meromorphic continuation 
from the domain $\Re\lambda>0$ to the Riemann surface 
with the branching points  $\pm im_n$ with $m_n\not=0$, $n\in\bar d$.
\end{cor}

Now we investigate the limit values of $M^{-1}(\lambda)$
at the imaginary axis $\lambda=ix$, $x\in\R$.
Without loss of generality, we assume that $d=2$ and $0<m_1<m_2$.
The other cases can be considered similarly.
The limit matrix
\be\la{a3}
M(ix+0)=\left(\ba{cc}
ix I&-I\\ \omega^2I-H(ix+0)&ix I \ea\right),\,\,\,\,x\in\R,
\ee
exists, and its entries are continuous functions of $x\in\R$,
smooth for $|x|<m_1$, $m_1<|x|<m_2$, and $|x|>m_2$.
\begin{lemma}
The limit matrix $M(ix+0)$ is invertible for $x\in\R$.
\end{lemma}
{\bf Proof}. (i) Let $|x|\le m_1$. Then the matrix
$(\omega^2-x^2)I-H(ix+0)$ is positive definite.
Indeed, for every $v\in\R^3\setminus\{0\}$, by the condition {\bf A1}
with $m_*=m_1$,
\beqn
v\cdot ((\omega^2-x^2)I-H(ix+0))v
&=&(\omega^2-x^2)|v|^2-
\sum\limits_{n=1}^d (2\pi)^{-3}
\int\frac{(k\cdot v)^2|\hat\rho_n(k)|^2\,dk}{k^2+m_n^2-x^2}
\nonumber\\
&\ge& (\omega^2-m_1^2)|v|^2-v\cdot Kv>0.
\nonumber
\eeqn
(ii) Let $m_1<|x|\le m_2$. Then $v\cdot\Im H(ix+0) v\not=0$
for every $v\in\R^3\setminus\{0\}$.
Indeed,
$$
\Im H_{ij}(ix+0)=\Im (2\pi)^{-3}
\int\frac{k_ik_j|\hat\rho_1(k)|^2\,dk}{k^2+m_1^2-(x-i0)^2}.
$$
For $\ve>0$, consider the function
$$
h_{ij}(ix+\ve)=\int\frac{k_ik_j|\hat\rho_1(k)|^2}{k^2+m_1^2-(x-i\ve)^2}\,dk,
\quad |x|>m_1.
$$
Denote $D_\ve(k)=k^2+m_1^2-(x-i\ve)^2$. For $|x|>m_1$,
$D_0(k)=0$ if $|k|=\sqrt{x^2-m_1^2}$.
We fix a small $\delta>0$ and introduce a cutoff function 
$\zeta\in C_0^\infty(\R^3)$
such that $\zeta(k)\ge0$, $\zeta(k)=1$ when $|D_0(k)|<\delta$
and $\zeta(k)=0$ when $|D_0(k)|\ge 2\delta$.
Note that $\Im h_{ij}(ix+0)=\Im h^\delta_{ij}(ix+0)$, where
$$
h^\delta_{ij}(ix+0)=\lim_{\ve\to0}\int\zeta(k)
\frac{k_ik_j|\hat\rho_1(k)|^2}{D_\ve(k)}\,dk.
$$  
Denote $a(k)=\sqrt{k^2+m_1^2}$. Assume that $x>0$. Since
$$
\frac1{D_\ve(k)}=\frac1{2a(k)(a(k)-x+i\ve)}+\frac1{2a(k)(a(k)+x-i\ve)},
$$
 $\Im h^\delta_{ij}(ix+0)=\Im h^\delta_{-}(ix+0)$, where
$$
h^\delta_{-}(ix+\ve):=\int\zeta(k)
\frac{k_ik_j|\hat\rho_1(k)|^2}{2a(k)(a(k)-x+i\ve)}\,dk.
$$
We rewrite $h^\delta_{-}(ix+\ve)$ as
$$
h^\delta_{-}(ix+\ve)=\int\frac{g(\alpha)}{\alpha+i\ve}\,d\alpha,\quad
g(\alpha)=\int\limits_{a(k)-x=\alpha}\zeta(k)
\frac{k_ik_j|\hat\rho_1(k)|^2}{2a(k)|\nabla a(k)|}\,dS.
$$
Hence $\Im h^\delta_{-}(ix+0)=-\pi g(0)$
by the Plemelj formula. Finally, for $x>m_1>0$,
$$
\Im h_{ij}(ix+0)=-\pi\int\limits_{|k|=\sqrt{x^2-m_1^2}}
\frac{k_ik_j|\hat\rho_1(k)|^2}{2|k|}\,dS.
$$ 
Hence, applying condition {\bf A3} we obtain that, for $m_1<|x|\le m_2$, 
\beqn\nonumber
v\cdot\Im H(ix+0)v=-{\rm sign}(x)\pi(2\pi)^{-3}
\int\limits_{|k|=\sqrt{x^2-m_1^2}}
\frac{(v\cdot k)^2|\hat\rho_1(k)|^2}{2|k|}\,dS\not=0.
\eeqn
(iii) Let $|x|>m_2$. In this case we find
\beqn\nonumber
v\cdot\Im H(ix+0)v=-{\rm sign}(x)\pi\sum\limits_{n=1}^2(2\pi)^{-3}
\int\limits_{|k|=\sqrt{x^2-m_n^2}}
\frac{(v\cdot k)^2|\hat\rho_n(k)|^2}{2|k|}\,dS\not=0.
\eeqn
In general case $d=1,2,\dots,$ we enumerate $m_1,\dots,m_d$
in the increasing order, $0\le m_1\le m_2\le\dots\le m_d$.
If $m_k\not=m_{k+1}$ and $m_k<|x|\le m_{k+1}$,  
\be\la{a2}
v\cdot\Im H(ix+0)v=-{\rm sign}(x)\pi \sum\limits_{n=1}^k
(2\pi)^{-3}\int\limits_{|k|=\sqrt{x^2-m_n^2}}
\frac{(v\cdot k)^2|\hat\rho_n(k)|^2}{2|k|}\,dS\not=0
\ee
by condition {\bf A3}.
For $|x|>m_d$, the formula (\ref{a2}) holds with $k=d$. 
\bo
\begin{remark}\la{r7.6}
We use condition {\bf A3} only in the estimate (\ref{a2}). Hence, instead of 
condition {\bf A3} it suffices to assume that 
for any $v\in\R^3\setminus \{0\}$ and $x>0$
$$
\sum\limits_{n:\,m_n<x}\Big(|k|^3
\int\limits_{|\theta|=1}
(v\cdot \theta)^2|\hat\rho_n(|k|\theta)|^2\,dS_{\theta}\Big)
\Big|_{|k|=\sqrt{x^2-m_n^2}}\not=0.
$$
\end{remark}
\begin{cor}
The matrix $M^{-1}(ix+0)$ is smooth w.r.t. $x\in\R$
outside the points $x=\pm im_l$ with $m_l\not=0$.
\end{cor}
\begin{lemma} \la{16.2}
(i)
The matrix $M^{-1}(ix+0)$ admits the following Puiseux expansion 
in a neighborhood of $\pm im_l$ ($m_l\not=0$): there exists an $\ve_\pm>0$
such that
\be\la{a5}
M^{-1}(ix+0)=\sum\limits_{k=0}^\infty c_k^{\pm}(x\mp m_l)^{k/2},\quad
|x\mp m_l|<\ve_\pm,\quad x\in\R.
\ee
(ii) There exists a matrix $R_0$ and a matrix-valued function $R_1(x)$
such that
\be\la{a6}
M^{-1}(ix+0)=\frac{1}{x}R_0+R_1(x),\quad |x|>\max_n m_n+1,\quad x\in\R,
\ee
where $|\pa_x^k R_1(x)|\le C_k/|x|^2$ for $|x|>\max_n m_n+1$, $x\in\R$,
 $k=0,1,\dots$.
\end{lemma}
{\bf Proof}.  
(i) Formula (\ref{a5}) follows from (\ref{gn}) and (\ref{M}). 

(ii) Let $f\in L^2(\R^3)$ with $\supp f\subset B_R$.
Then (see formula (16.7) of \cite{IKV})
$$
\Vert \pa^k_x [-\Delta+m_n^2+(ix+0)^2]^{-1} f\Vert_{L^2(B_R)}\le
\frac{C_k(R)}{|x|}\Vert f\Vert _{L^2(B_R)},\quad |x|\ge m_n+1,
$$
for every $R>0$. Therefore (see (\ref{4.9'})) we obtain that
 $|\pa_x^k H_{ij}(ix+0)|\le C_k/|x|$ for
$|x|>\max_n m_n+1$. Then formula (\ref{a3}) implies (\ref{a6}). \bo
\medskip\\
{\bf Proof of Theorem \ref{Appendix}}.
Applying (\ref{7.5}) we obtain that
\be\la{a7}
\left(\ba{c} q(t)\\ p(t)\ea\right)=
\frac1{2\pi}\int\limits_{-\infty}^{+\infty}
e^{ixt}M^{-1}(ix+0)\left(\ba{c} q^0\\  \bar p^0(ix+0)\ea\right)\,dx.
\ee
Without loss of generality, let us assume that
$0<m_1<m_2<\dots<m_d$.
We split the Fourier integral (\ref{a7}) into $d+1$ terms by using 
the partition of unity $\zeta_0(x)+\dots+\zeta_d(x)=1$,
$x\in\R$:
\beqn
\left(\ba{c} q(t)\\ p(t)\ea\right)&=&
\frac1{2\pi}\int\limits_{-\infty}^{+\infty}
e^{ixt}(\zeta_0(x)+\dots+\zeta_d(x))M^{-1}(ix+0)
\left(\ba{c} q^0\\ \bar p^0\ea\right)\,dx\nonumber\\
&=&I_0(t)+\dots+I_d(t),
\nonumber\eeqn
where
the functions $\zeta_k(x)\in C^\infty(\R)$ are supported by
\beqn
&&\supp\zeta_0\subset\{x\in\R:\,|x|>m_d+1 \,\mbox{or }\,|x|<m_{1}/2\},
\nonumber\\
&&\supp\zeta_1\subset\{x\in\R:\,m_{1}/3<|x|<(m_1+m_2)/2\},\nonumber\\
&&\supp\zeta_2\subset\{x\in\R:\,m_2-2(m_2-m_1)/3<|x|<(m_2+m_3)/2\},\dots,\nonumber\\
&&\supp \zeta_d\subset\{x\in\R:\,m_d-2(m_d-m_{d-1})/3<|x|<m_{d}+2\}.
\nonumber\eeqn
Then
 (i) the function $I_0(t)\in C[0,+\infty)$ decays faster than any power of $t$ 
due to Lemma \ref{16.2}, and
(ii) the functions $I_k(t)\in C^\infty(\R)$, $k=1,\dots,d$, 
decay like $(1+|t|)^{-3/2}$
by virtue to (\ref{a5}). Theorem \ref{Appendix} is proved.\bo
\begin{cor}
Let $m\not=0$ and conditions {\bf A1}--{\bf A3} hold. Then
\be\la{7.10}
|{\cal N}^{(j)}(t)|\le C(1+|t|)^{-3/2},\,\,\,j=0,1,
\ee
where ${\cal N}(t)$ is defined in (\ref{4.15}) and (\ref{A}).
This bound can be proved by similar way as Theorem \ref{Appendix}.
\end{cor}


\end{document}